\documentclass[aps,prb,twocolumn,showpacs,superscriptaddress,10p,longbibliography]{revtex4-2}

\usepackage{amssymb}
\usepackage{graphicx}
\usepackage[colorlinks = true, urlcolor = blue, linkcolor = blue, citecolor = blue]{hyperref}
\usepackage[sort&compress]{natbib}
\usepackage{scalerel}
\usepackage[normalem]{ulem}
\usepackage[dvipsnames]{xcolor}
\usepackage{bm}
\usepackage{amsmath}
\usepackage{subfigure}
\usepackage{amsfonts}
\usepackage{mathtools}
\usepackage{physics}
\usepackage{dsfont}
\usepackage{float}
\usepackage{lmodern}
\usepackage[many]{tcolorbox}
\usepackage{empheq}
\usepackage{soul}
\usepackage[left]{lineno}

\newcommand{\mh}{\mathcal{H}}

\newcommand{\mcp}{\mathcal{P}}
\newcommand{\mct}{\mathcal{T}}
\newcommand{\mpt}{\mathcal{PT}}
\newcommand{\mr}{\mathcal{R}}

\newcommand{\mF}{\mathcal{F}}
\newcommand{\vmE}{\vec{\mathcal{E}}}
\newcommand{\vn}{\vec{n}}
\newcommand{\vl}{\vec{l}}

\newcommand{\vk}{\vec{k}}

\usepackage{tikz}
\definecolor{lime}{HTML}{A6CE39}
\DeclareRobustCommand{\orcidicon}{%
    \begin{tikzpicture}
    \draw[lime, fill=lime] (0,0) 
    circle [radius=0.13] 
    node[white] {{\fontfamily{qag}\selectfont\tiny ID}};
    \draw[white, fill=white] (-0.0625,0.095) 
    circle [radius=0.007];
    \end{tikzpicture}
    \hspace{-2mm}}
    
\newcommand{\orcidAM}{\href{https://orcid.org/0000-0002-7307-5922}{\orcidicon}}
\newcommand{\orcidAA}{\href{https://orcid.org/0000-0002-3033-0452}{\orcidicon}}

\newcommand{\pcsadd}{Center for Theoretical Physics of Complex Systems, \href{https://ror.org/00y0zf565}{Institute for Basic Science} (IBS), Daejeon 34126, Republic of Korea}
\newcommand{\ustadd}{Basic Science Program, \href{https://ror.org/000qzf213}{Korea University of Science and Technology} (UST), Daejeon 34113, Korea}

\begin{document}

\title{Flat bands in tight-binding lattices with anisotropic potentials}

\author{Arindam Mallick\orcidAM}
\email{jhunuguma@gmail.com}
\affiliation{Instytut Fizyki Teoretycznej, \href{https://ror.org/03bqmcz70}{Uniwersytet Jagielloński}, \L{}ojasiewicza 11, 30-348 Kraków, Poland}
\affiliation{\pcsadd}

\author{Alexei Andreanov\orcidAA}
\email{aalexei@ibs.re.kr}
\affiliation{\pcsadd}
\affiliation{\ustadd}

\date{\today}

\begin{abstract}
We consider tight-binding models on Bravais lattices with anisotropic onsite potentials that vary along a given
direction and are constant along the transverse one. Inspired by our previous work on flat bands in anti-$\mathcal{PT}$ symmetric Hamiltonians [Mallick {\it et al.}, \href{https://doi.org/10.1103/PhysRevA.105.L021305}{Phys.~Rev.~A {\bf 105}, L021305 (2022)}], we construct an anti-$\mathcal{PT}$ symmetric Hamiltonians with an $E = 0$ flat band by tuning the hoppings and the shapes of potentials. This construction is illustrated for the square lattice with bounded and unbounded potentials. Unlike flat bands in short-ranged
translationally invariant Hamiltonians, we conjecture that the considered $E = 0$ flat bands do not host compact
localized states. Instead the flat-band eigenstates exhibit a localization transition along the potential direction
upon increasing the potential strength for bounded potentials. For unbounded potentials flat-band eigenstates are
always localized irrespective of the potential strength.
\end{abstract}

\maketitle

\section{Introduction}

Flat bands are dispersionless bands in tight-binding Hamiltonians~\cite{leykam2018artificial, yang2012topological}, which are macroscopically degenerate, typically host compact localized states (CLSs), and exhibit localization without disorder \cite{maimaiti2020thesis, maimaiti2021flatband}.
Flat bands attract a lot of attention due to their extreme sensitivity to perturbations and emergent novel phases of matter with applications in various fields ranging from condensed matter systems to quantum technologies, e.g., high temperature  superconductivity~\cite{kobayashi2016superconductivity,iglovikov2014superconducting,kopnin2011high,mondaini2018pairing,volovik2018graphite},
subdimensional localization~\cite{mukherjee2021minimal, mallick2022correlated}, quantum chaos~\cite{kuno2020flat_qs, kuno2021multiple}, quantum hardware~\cite{lai2016geometry,rontgen2019quantum, yang2023photonic}, etc.

In translation invariant lattices, the construction of flat bands and their compact localized eigenstates has been studied extensively in the past years~\cite{maimaiti2017compact, leykam2018artificial, maimaiti2019universal, rhim2019classification, zhang2020compact, roentgen2018compact, travkin2017compact, graf2021designing, danieli2024flat}. However flat bands can also exist in nontranslationally invariance systems like quasicrystals~\cite{nori1990angular,dayroberts2020nature,ha2021macroscopically}.
Another example is a tight-binding Hamiltonian on a \(d\)-dimensional Bravais lattice in the presence of a uniform Wannier-Stark field (linear potential).
The field partially destroys translational invariance and all-bands-flat spectrum emerges \cite{mallick2021wannier}. These flat bands do not host CLSs, but rather noncompact superexponentially localized eigenstates. The Wannier-Stark flat bands were also studied in non-Bravais lattice settings~\cite{kolovsky2018topological, mallick2022antipt}. An interesting open issue is the existence of flat bands in other classes of nontranslationally invariant Hamiltonians and localization properties of their eigenstates.

In this paper, we propose Hamiltonians on Bravais lattices with an on-site potential that are antisymmetric under the joint action of reflection \(\mcp\) and time reversal \(\mct\).
The on-site potential varies along a given lattice direction and is constant in all the transverse directions.
Such Hamiltonians have a single \(E=0\) flat band unlike the dc field case~\cite{mallick2021wannier}, where all bands are flat. Our flat-band construction is valid for all
Bravais lattices and arbitrary range of hopping as well as any
on-site potential, so long as it varies along a direction that is a
lattice vector.

We provide several examples for the tight-binding Hamiltonians on the square lattice with bounded potentials:
quasi-periodic Aubry-Andr\'e (AA)-like potential~\cite{aubry1980analyticity}, an inverse trigonometric potential, and an unbounded potential. In contrast to the conventional flat bands which feature compact localized states and do not support delocalized eigenstates,
here, by analyzing Bloch Hamiltonian and inverse participation ratio for flat-band eigenstates, we demonstrate that a localization-delocalization transition along the field can happen.
This transition is consistent with the observations in Refs.~\cite{rossignolo2019localization, antonio2022coexistence}.
For the example of an unbounded polynomial potential, we find that the eigenstates of the flat band are always localized.

The article is organized as follows:
In Sec.~\ref{sec:model}, we define the generic Hamiltonian with anisotropic on-site potential.
In Sec.~\ref{sec:anti_symmtry}, we define the antisymmetry operator and analyze possible constraint on a tight-binding Hamiltonian with arbitrary long-range hopping in simple square lattices---which supports an \(E = 0\) flat band---and we provide the abstract existential proof of the flat band in a general \(d\)-dimensional Bravais lattice in Appendix~\ref{app:general_lattice}.
In Sec.~\ref{sec:numerical}, we provide numerical evidences of the flat band and possible localization-delocalization transition with both bounded and unbounded potentials.
We concluded in Sec.~\ref{sec:concl}.

\section{Model with an anisotropic potential}
\label{sec:model}

We consider a square lattice with sites labeled by an integer vector \(\vec{n} = (n_1,n_2) \in \mathbb{Z}^2\) and a tight-binding Hamiltonian with an on-site potential:
\begin{align}
  \mh = \sum_{\vn} \left[V(n_1, n_2) \ketbra{\vn}{\vn} - \sum_{\vl} t(\vl)\ketbra{\vn}{\vn + \vl}\right].
  \label{eq:main_hamil}
\end{align}
Here \(\vl = (l_1, l_2)\) is the lattice vector connecting lattice sites \(\vn,\vn + \vl\).
The Hermiticity of the Hamiltonian enforces \([t(-\vl)]^* = t(\vl)\).

\begin{figure}[]
  \includegraphics[width = 0.75\columnwidth]{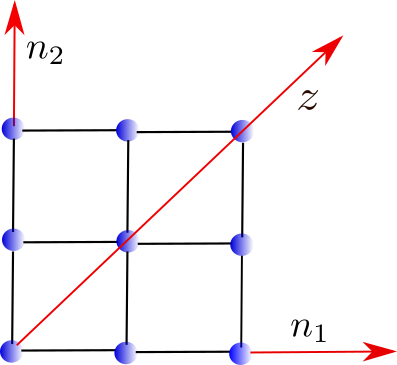}
  \caption{Blue spheres indicate lattice sites, black solid lines are hopping connections, red arrowed lines are for the direction \(z\) and lattice basis vectors \(n_1, n_2\).
    Square hopping network with tilted field: the field is along the diagonal with respect to the unit cell basis vectors.}
  \label{fig:1_1_square}
\end{figure}

We assume that the potential \(V(n_1, n_2)\) varies along a lattice vector \(\vec{z}\) and is constant in the transverse direction \(\vec{w}\), which is also a lattice vector for the square lattice. The present setting can be considered as a generalization of the dc field case, where the dc field was applied along the direction \(\vec{z}\), that we considered previously in Ref.~\cite{mallick2021wannier}.
Since both \(\vec{z}, \vec{w}\) are lattice vectors, we can express them as follows:
\begin{align}
  z = \alpha_1 n_1 + \alpha_2 n_2, \quad w = \alpha_2 n_1 - \alpha_1 n_2,
\end{align}
where \(|\alpha_{1}|\) and \(|\alpha_2|\) are either mutually prime numbers or tuples \(\{(0, \pm 1), (\pm 1, 0)\}\)~\footnote{\((\alpha_{1}, \alpha_2)\) are related to the DC field direction for Wannier-Stark system~\cite{mallick2021wannier}}.
In what follows, we use either the tuple \((\alpha_1, \alpha_2)\) or the decomposition of \(z\) to denote the direction in which the potential varies in space.
Thanks to the B\'ezout's identity, both \(z\) and \(w\) are ensured to take integer values only.
However, there are constraints on the allowed values of \(w\) depending on the value of \(z\) and vice versa, except for the simple cases when one of \(\alpha_{1,2}\) is zero~\cite{mallick2021wannier}.
For example, if the potential changes only along the main diagonal of the square lattice, \((1, 1)\), as shown in Fig.~\ref{fig:1_1_square}, then \(z = n_1 + n_2, w = n_1 - n_2 = z - 2 n_2\) and \(w\) can only take even (odd) values for even (odd) \(z\).
Therefore, it is more convenient to work with \((z,\eta)\), \(\eta \equiv n_2\) rather than \((z,w)\) in this case.
The proper way of defining \(\eta\) was defined in Ref.~\cite{mallick2021wannier} for a generic 2D Bravais lattice and generic tuples \((\alpha_1, \alpha_2)\).

The translation invariance of the Hamiltonian along \(w\) implies the invariance in \(\eta\).
Therefore, the Hamiltonian~\eqref{eq:main_hamil} can be block diagonalized using the Fourier transform with an associated momentum \(k\):
\begin{align}
  & \mh = \sum_{k} \mh(k) \otimes \ketbra{k}{k}, \notag \\
  & \mh(k) = \sum_{z} V(z) \ketbra{z}{z} - \sum_{z,\vl} t(\vl) e^{i k l_\eta} \ketbra{z}{z+l_z},
  \label{eq:main_hamil_k}
\end{align}
where \(l_z = \alpha_1 l_1 + \alpha_2 l_2\) is the hopping step in the \(z\) direction, and \(l_\eta\) is the step of the \(z\)-independent coordinate \(\eta\) along the hopping direction.
The value of \(l_\eta\) is a linear function of \(\vl\); its exact form depends on the details of the Hamiltonian~\cite{mallick2021wannier} and is provided below for several examples.
We have also assumed either periodic or infinite system along \(\vec{w}\) to properly define the momentum \(k\).

The other Bravais lattices are discussed in Appendix~\ref{app:general_lattice}.

\section{Antisymmetry-induced flat band}
\label{sec:anti_symmtry}

We define a time-reversal operator \(\mct: \ket{\psi} \mapsto \ket{\psi^*}\) (with respect to the position basis) and a parity operator \(\mcp\),
\begin{align}
  \label{eq:parity}
  \mcp = \sum_{\vn} (-1)^{z} \ket{-\vn}\bra{\vn},
\end{align}
which reflects the lattice site vector \(\vn\) with respect to the origin \(\vn=\vec{0}\) and adds a \(z\)-dependent phase. We require an antisymmetry of the full Hamiltonian~\eqref{eq:main_hamil} under the joint action of \(\mcp\) and \(\mct\): \(\mr = \mct \cdot \mcp\):
\begin{align}
  \mr \cdot \mh \cdot \mr^{-1} = -\mh\;.
  \label{eq:anti_symmtry_con}
\end{align}
This requires the potential \(V\) to be odd with respect to the reflection of \(\vec{z}\)~\eqref{eq:parity},
\begin{align}
  \label{eq:odd_pot}
  V(-z) = -V(z),
\end{align}
and imposes a nontrivial constraint on the hoppings:
\begin{align}
  \label{eq:hopping_restrict}
  [t(-\vl)]^* = -(-1)^{l_z} t(\vl)\;.
\end{align}
The Hermiticity condition, \([t(-\vl)]^* = t(\vl)\), implies that only odd \(l_z\) are allowed.
As a consequence, \(l_z \neq 0\) and no equipotential hopping is allowed---similarly to the Wannier-Stark flat-band case~\cite{mallick2021wannier}.
The absence of equipotential hopping holds for any Bravais lattice model---see Eq.~\eqref{appeq:hopp_cond} in Appendix~\ref{app:general_lattice}.
There might exist other possible choices of the \(R\).

Now we observe that the momentum \(\ket{k} \mapsto \ket{-k}\) under the action of either \(\mcp\) or \(\mct\), but it is invariant under the joint action \(\mct \cdot \mcp\).
Therefore, the Hamiltonian \(\mh\) is anti-symmetric under the action of \(\mct \cdot \mcp\) for each momentum \(k\) independently:
\begin{align}
  \mr_k \cdot \mh(k)  = -\mh(k) \cdot \mr_k~, ~ \mr_k = \mel{k}{\mct \cdot \mcp}{k}\;.
\end{align}
Therefore, the eigenvalues of \(\mh\) come in pairs \((E(k), -E(k))\) with corresponding eigenvectors \((\ket{\psi(k)}\), \(\mel{k}{\mct \cdot \mcp}{k}\ket{\psi(k)})\) for each momentum \(k\).
For an odd number of available \(z\) values, the Hamiltonian \(\mh(k)\) has an odd number of eigenvalues for each \(k\).
Consequently, there is necessarily zero eigenvalues, \(E(k) = 0\), for all momenta \(k\), i.e., a flat band, which is located at the middle of the spectrum.
This is similar to the case of chiral flat bands~\cite{kolovsky2018topological} and the requirement of the odd number of sublattices for the existence of anti-\(\mathcal{PT}\) symmetric flat bands for non-Bravais lattices~\cite{mallick2022antipt}.
For a finite system of size \(L\) along the \(z\) direction, this implies an odd number of equipotential lines (value of \(z\) is fixed on each of the lines): \(z \in [-(L-1)/2,(L-1)/2]\).
This implies an odd number of energy bands \(E(k)\).

\subsection{Two examples of the potential \(V\)}
\label{sec:tilted_square}

We provide two example settings on the square lattice that satisfy the above requirements.

One example is shown in Fig.~\ref{fig:1_1_square}: the potential only varies along the main diagonal of the square lattice, \(z = n_1+n_2\).
For the nearest-neighbor hopping, we have \((l_1, l_2) \in  \{(0, \pm 1), (\pm 1, 0)\}\).
Therefore, \(l_z \in \{1, -1\}\) and there is no equipotential hopping [Eq.~\eqref{eq:hopping_restrict}], i.e., no hopping is present along the anti-diagonal, \(w = n_1 - n_2\) in the absence of hopping along \(z\).
The corresponding Hamiltonian reads
\begin{align}
  & \mh_{1,1} = \sum_{n_1, n_2} \bigg[ V(n_1 + n_2) \ketbra{n_1, n_2}{n_1, n_2} \notag \\
  & - \sum_{l = \pm 1} \ketbra{n_1, n_2}{n_1 + l, n_2}  + \ketbra{n_1, n_2}{n_1, n_2 + l}\bigg]\;.
  \label{eq:2D_(1_1)}
\end{align}
In this case, the antidiagonal coordinate \(w\) is a function of \(z\) and contains a \(z\)-independent part.
We define \(\eta := n_2\) as the independent coordinate~\cite{mallick2021wannier}:
\begin{align}
  z = n_1 + n_2, \quad w = n_1 - n_2 = z - 2\eta\;.
  \label{eq:coordinate_def_11}
\end{align}
The Hamiltonian \(\mh_{1,1}\) reads in terms of the \((z, \eta)\) coordinates,
\begin{align}
  & \mh_{1,1} = \sum_{z, \eta} \bigg[ V(z) \ketbra{z, \eta}{z, \eta} \notag\\
  & - \sum_{l = \pm 1} \ketbra{z, \eta}{z + l, \eta}  + \ketbra{z, \eta}{z + l, \eta + l} \bigg], 
  \label{eq:2D_(1_1)_eta}
\end{align}
and is partially diagonalized by the Fourier transform with respect to coordinate \(\eta\) for fixed \(z\):
\begin{gather}
  \mh_{1,1}(k) = \sum_{z} \left[ V(z) \ketbra{z}{z} - \sum_{l = \pm 1} (1 + e^{ikl})\ketbra{z}{z + l} \right] \;.
  \label{eq:2D_(1_1)_k}
\end{gather}
Note that for \(k l = \pm \pi\), the hopping part vanishes, leading to compact localized structure of the eigenstates for all eigenvalues.

\begin{figure}[]
  \includegraphics[width = 0.9\columnwidth]{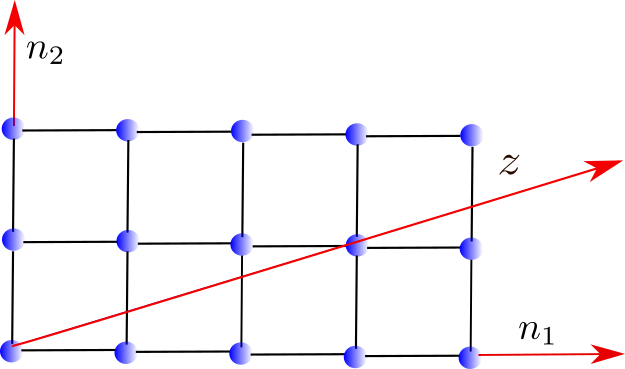}
  \caption{
    Blue spheres indicate the square lattice sites. 
    The indices are \((n_1,n_2)\) with the corresponding basis vector directions indicated by the red arrow.
    The red arrow with the label \(z\) indicates the \(z\) direction.
  }
  \label{fig:3_1_square}
\end{figure}

For the more tilted case, \((3, 1)\), shown in Fig.~\ref{fig:3_1_square}, the rotated coordinates read
\begin{align} 
  z &= 3n_1+n_2, \notag \\
  \label{eq:coordinate_def_31}
  w &= n_1 - 3 n_2 = n_1 - 3(z - 3n_1) = -3z + 10 n_1, 
\end{align}
and we choose \(n_1 \equiv \eta\).
The square lattice Hamiltonian \eqref{eq:main_hamil} with the usual nearest-neighbor hopping becomes 
\begin{align}
  \mh_{3,1} = & \sum_{z, \eta} \bigg[V(z) \ketbra{z, \eta}{z, \eta} \notag \\
  & - \sum_{l = \pm 1} \ketbra{z, \eta}{z + 3l, \eta + l} +  \ketbra{z, \eta}{z + l, \eta}\bigg]\;.
  \label{eq:2D_(3_1)_k}
\end{align}
In the momentum space (with respect to \(\eta\)), the Hamiltonian  reads
\begin{align}
  \label{eq:(3_1)_ham_square_k}
  \mh_{3,1}(k) = \sum_{z} &\bigg[V(z) \ketbra{z}{z} \notag\\
  & - \sum_{l = \pm 1} e^{ikl}\ketbra{z}{z + 3l} + \ketbra{z}{z + l}\bigg]\;.
\end{align}

In both cases, if we choose odd potential, \(V(-z) = -V(z)\), then the Hamiltonian is anti-symmetric under the action of \(\mct \cdot \mcp\). 
Other directions of the potential are discussed in Appendix~\ref{app:general_lattice}.

\section{Numerical results}
\label{sec:numerical}

We now verify the above construction, Eqs.~\eqref{eq:2D_(1_1)_k} and \eqref{eq:(3_1)_ham_square_k}, for several example potentials.
We consider quasi-periodic potential, which is easily realizable in state-of-the-art experimental devices, and trigonometric and  polynomial potentials \cite{zakrzewski2007tight}.
The Hamiltonians are defined on the square lattice wrapped around a cylinder in the \((z, k)\) space, assuming periodicity along the \(k\) direction.
Along the \(z\) direction (the potential), we assume the cylinder large but finite.
We denote the size along \(z\) as \(L\) and refer to it as the system size in what follows.
The momentum-space Hamiltonians given by Eqs.~\eqref{eq:2D_(1_1)_k} and~\eqref{eq:(3_1)_ham_square_k} are diagonalized numerically.

The macroscopic degeneracy of the flat-band eigenstates \(\ket{\psi_{E = 0}(k)}\) is a challenge: any linear combination of eigenstates is also an eigenstate.
Translationally invariant Hamiltonians with a flat band host CLS~\cite{read2017compactly}, which provide a convenient basis for analysis~\cite{danieli2024flat}.
However, we conjecture that flat bands under consideration do not have CLS, as could potentially be shown by extending the same argument used for the Wannier-Stark flat bands~\cite{mallick2021wannier}.
We choose the \(z,k\) basis, that is natural for the model, to study the flat-band eigenstates and their dependence on the potential strength \(\lambda\). This implies, in particular, that eigenstates in the direction transverse to \(z\) are always delocalized for this choice of flatband eigenstates. To quantify localization of the states along \(z\), we compute the participation entropy:
\begin{align}
  \label{eq:ipr_ent}
  \mathcal{S}(k,\lambda) = -\ln\left(\sum_z |\psi_{E = 0}(z,k)|^4\right)\;.
\end{align}
\(\mathcal{S}(k,\lambda) = \ln(L)\) for fully delocalized state of size \(L\) along the \(z\) direction, and \(\mathcal{S}(k,\lambda) = 0\) for a state occupying a single site.
Therefore \(\mathcal{S}(k,\lambda)/\ln(L)\) is a monotonically increasing function of localization volume of a state.

\subsection{Aubry-Andr\'e potential}
\label{sec:aa_flat}

\begin{figure}
  \subfigure[]{\includegraphics[width = 0.45\columnwidth]{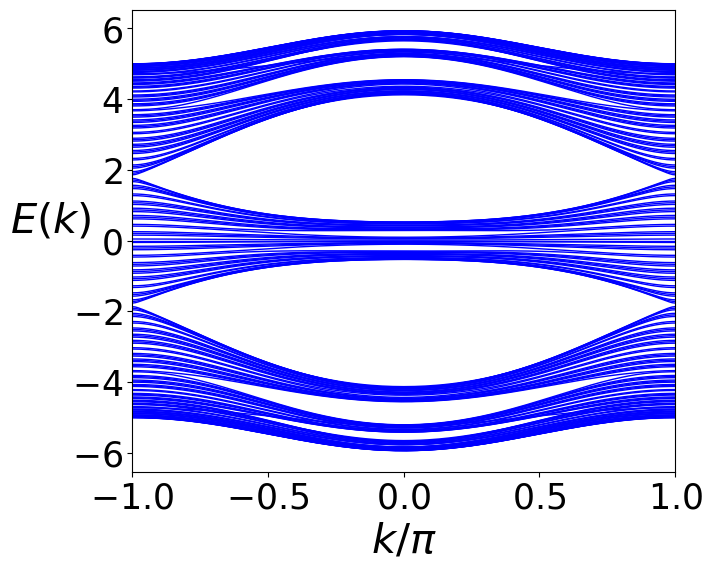}}
  \subfigure[]{\includegraphics[width = 0.47\columnwidth]{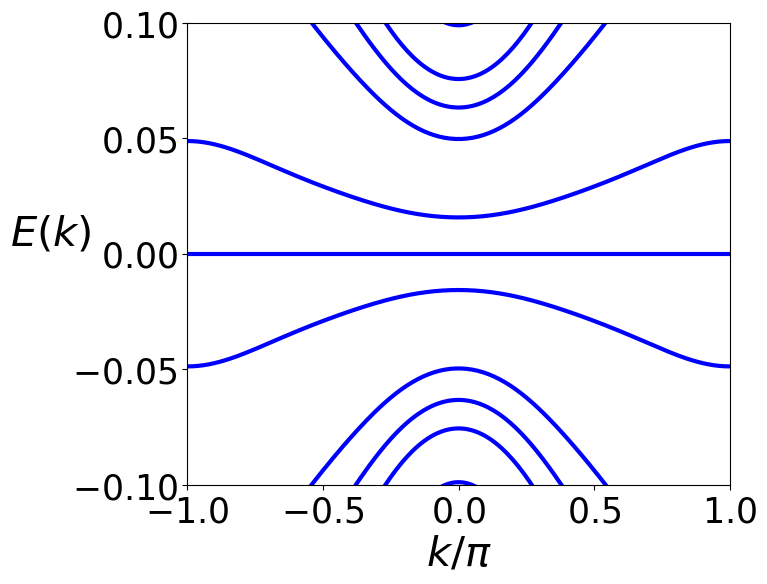}}
  \subfigure[]{\includegraphics[width = 0.45\columnwidth]{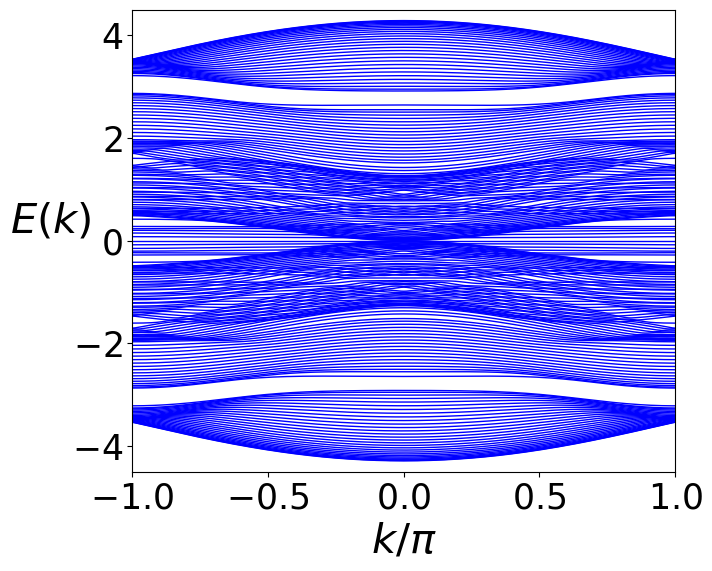}}
  \subfigure[]{\includegraphics[width = 0.47\columnwidth]{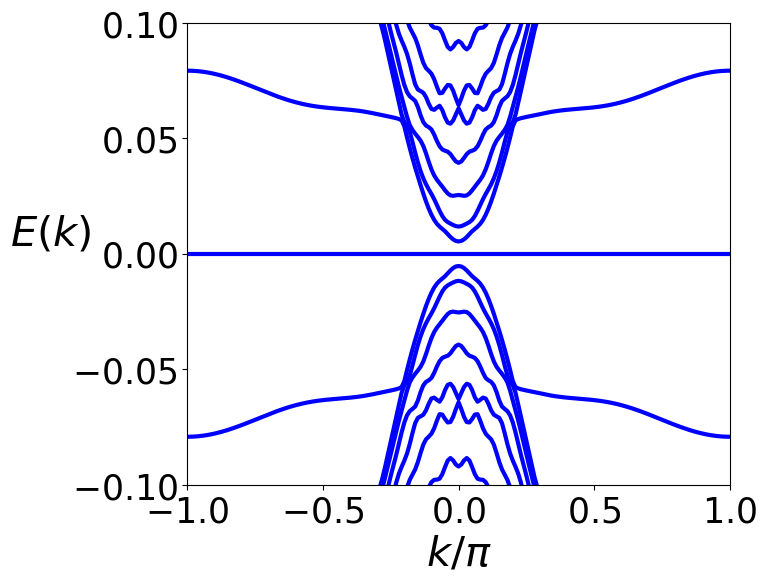}}
  \caption{
    (a)-(b) Band structure containing flat band at \(E = 0\) for the Hamiltonian in Eq.~\eqref{eq:2D_(1_1)_eta} with \(\lambda = 5\).
    (c)-(d) Band structure containing flat band at \(E = 0\) for the Hamiltonian in Eq.~\eqref{eq:2D_(3_1)_k}  with \(\lambda = 1.6\).
    (b), (d) The spectra shown in subfigures (a) and (c), respectively, magnified around the \(E=0\) flat band.
    \(201\) lattice sites along \(z\) are considered with open boundary condition.
    The reference (center of reflection for parity operation) site \(\vn = \vec{0}\) is set at the center of the lattice.
    \(\beta = \left(\sqrt{5} + 1\right)/2\).
  }
  \label{fig:aa_flat2}
\end{figure}

We consider the AA potential of strength \(\lambda\) and irrational parameter \(\beta\),
\begin{align}
  \label{eq:1D_sin_aa}
  V(z) = \lambda \sin(2\pi\beta z),
\end{align}
for the two Hamiltonians in Eqs.~\eqref{eq:2D_(1_1)_k} and~\eqref{eq:(3_1)_ham_square_k} varying along directions \((1, 1): z = n_1 + n_2\) and \((3, 1): z = 3 n_1 + n_2\), respectively.
Using exact diagonalization, we computed the spectrum for potential strengths \(\lambda = 5\) and \(\lambda = 1.6\) for directions (1,1) and (3,1), respectively, \(\beta = (1+\sqrt{5})/2\), and system size \(L = 201\) with open boundary condition along the \(z\) direction as shown in Fig.~\ref{fig:aa_flat2}.
The spectrum is symmetric with respect to \(E=0\) and a flat band is present at \(E = 0\) due to the choice of the odd system size \(L\).

\begin{figure}
  \subfigure[]{\includegraphics[width = 0.45\columnwidth]{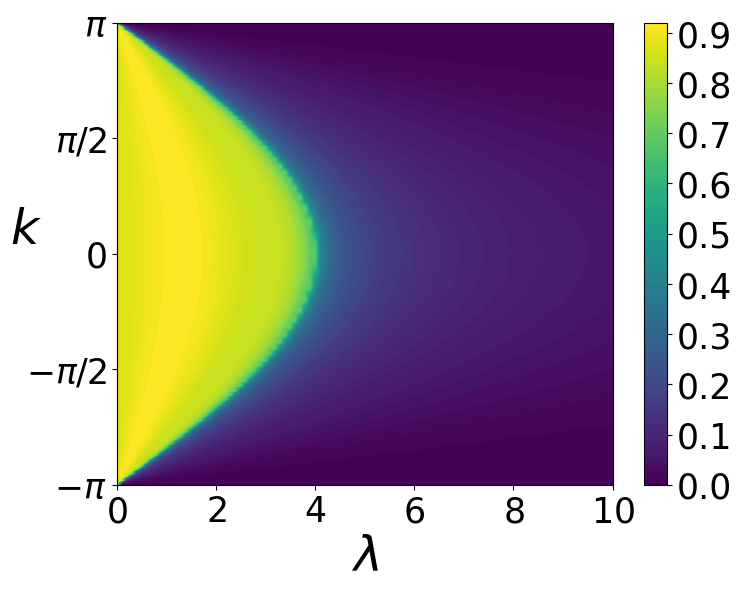}}
  \subfigure[]{\includegraphics[width = 0.47\columnwidth]{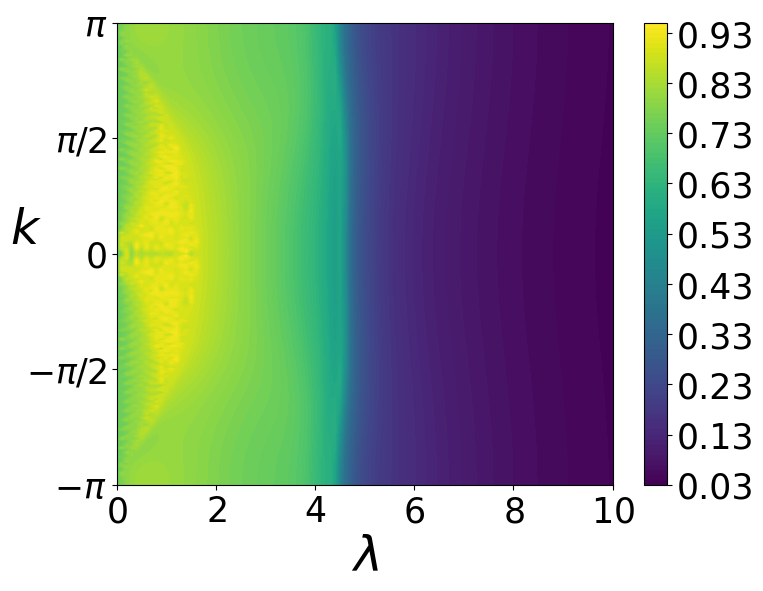}}
  \caption{
    \(\mathcal{S}(k,\lambda)/\ln(L)\) as a function of \(k, \lambda\) for the \(E = 0\) flat-band eigenstates for quasi-periodic potential \eqref{eq:1D_sin_aa}.
    System size \(L = 201\) along the field.
    (a) and (b) correspond to the potential directions as in Figs.~\ref{fig:1_1_square} and \ref{fig:3_1_square}, respectively.
  }
  \label{fig:pr_flat}
\end{figure}

\begin{figure} 
  \subfigure[]{\includegraphics[width = 0.23\textwidth]{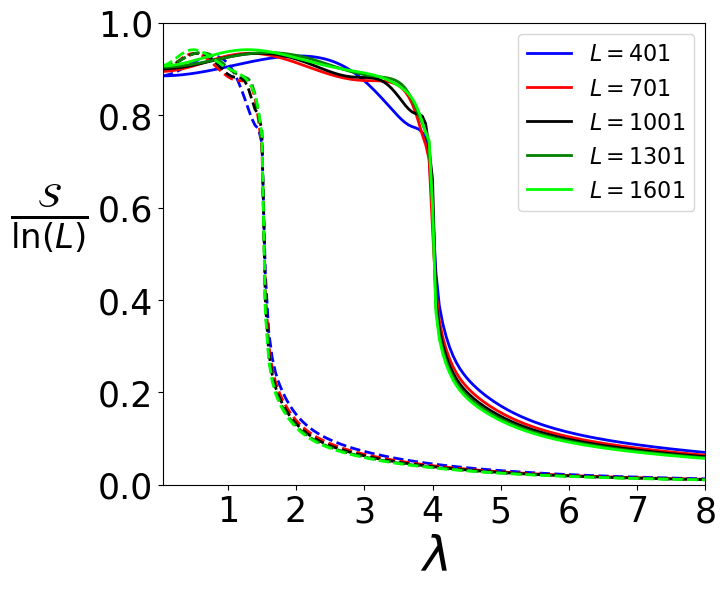}}
  \subfigure[]{\includegraphics[width = 0.23\textwidth]{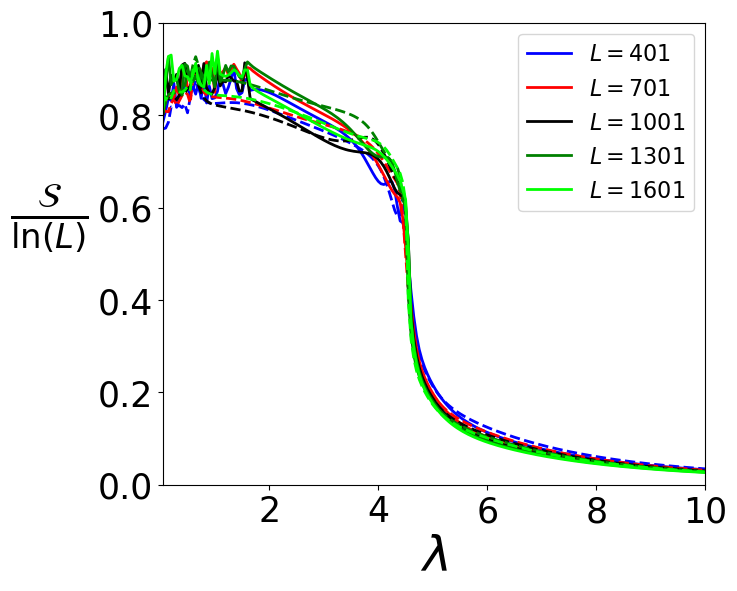}}
  \caption{Participation entropy in units of \(\ln(L)\) as a function of the potential strength \(\lambda\)~\eqref{eq:1D_sin_aa} for different system sizes \(L\) along the z-direction.
    Potential directions (a) \((1, 1), z = n_1 + n_2\) and (b) \((3, 1), z = 3n_1 + n_2\). Solid lines correspond to momentum \(k = 0\) while dashed lines correspond to \(k = 3\pi/4\).
    Note that for the direction (3, 1), the curves for the two momenta are barely distinguishable.}
  \label{fig:pr_scaling_aa}
\end{figure}

\begin{figure}
  \includegraphics[width = 0.45\textwidth]{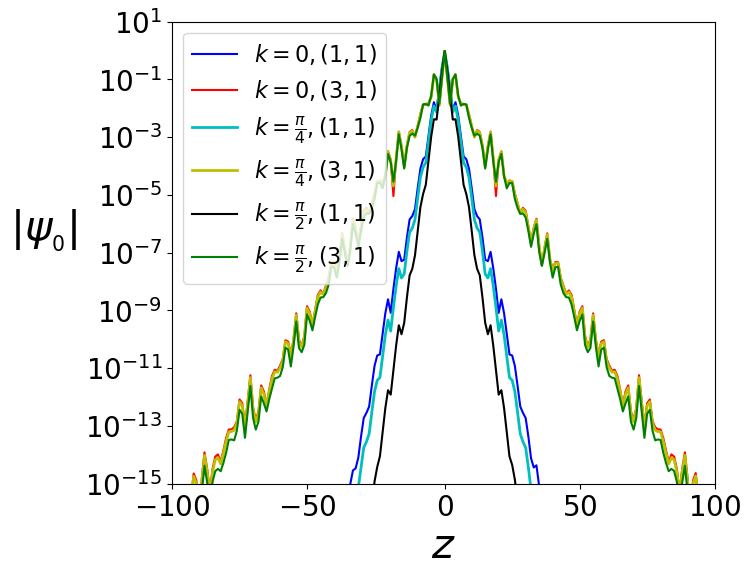}
  \caption{
    Profiles of the flat-band eigenstates as functions of \(z\) for different momenta and different orientation of the potential at fixed \(\lambda = 10\) \eqref{eq:1D_sin_aa}, where all \(E=0\) are localized along \(z\).
    The direction \((\alpha_1, \alpha_2) = (3,1)\) leads to larger localization volume than that for the direction \((\alpha_1, \alpha_2) = (1,1)\) for the same set of parameters.
  }
  \label{fig:aa_loc}
\end{figure}

The Hamiltonian \(\mh\) is diagonalized for a discrete set of \(k\) and several values of \(\lambda\), the participation entropy~\eqref{eq:ipr_ent} is then computed from the set of eigenvalues and eigenvectors \(\{E(k), \ket{\psi(k)}\}\).
The violet (dark) regions in Fig.~\ref{fig:pr_flat} correspond to the flat-band eigenstates localized along the field while yellow/green (bright) regions correspond to eigenstates delocalized along the \(z\) direction.
We observe a sharp, \(k\)-dependent transition from delocalized (yellow/green) to localized (violet) eigenstates with increasing potential strength \(\lambda\).
Similar results are obtained for other system sizes \(L\), as shown in Fig.~\ref{fig:pr_scaling_aa}.
The transition for the direction \((1, 1)\), i.e., \(z=n_1+n_2\) matches well with the analytical prediction~\cite{rossignolo2019localization}, up to additional phase factors (see Appendix~\ref{app:transition_deriv}).
\begin{align}
  \lambda = 4\left|\cos\left(\frac{k}{2}\right)\right|\;.
\end{align}
The transition is further confirmed by the finite size scaling of the participation entropy shown in Fig.~\ref{fig:pr_scaling_aa}.
Figure~\ref{fig:aa_loc} shows localized profiles of several \(E=0\) eigenstates for momenta \(k=0,\pi/4,\pi/2\) and the potential directions \((1, 1), (3, 1)\).
Profiles get more localized for larger momenta \(k\), and also for direction \((1, 1)\) compared to \((3, 1)\), while all the other parameters are the same.

We have considered a finite irrational parameter \(\beta=(\sqrt{5} + 1)/2\).
Different behaviors, e.g., pronounced intermediate localization completely hiding exponential decay, might emerge upon decreasing the value of \(\beta\), extending the results of Ref.~\cite{mallick2023intermediate}.

\subsection{Arctan potential}
\label{sec:frac_flat}

\begin{figure}
\subfigure[]{\includegraphics[width = 0.45\columnwidth]{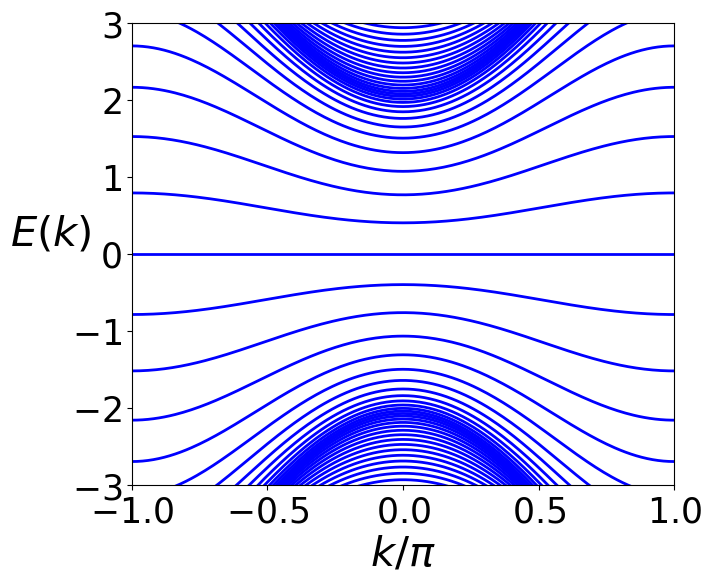}}
\subfigure[]{\includegraphics[width = 0.45\columnwidth]{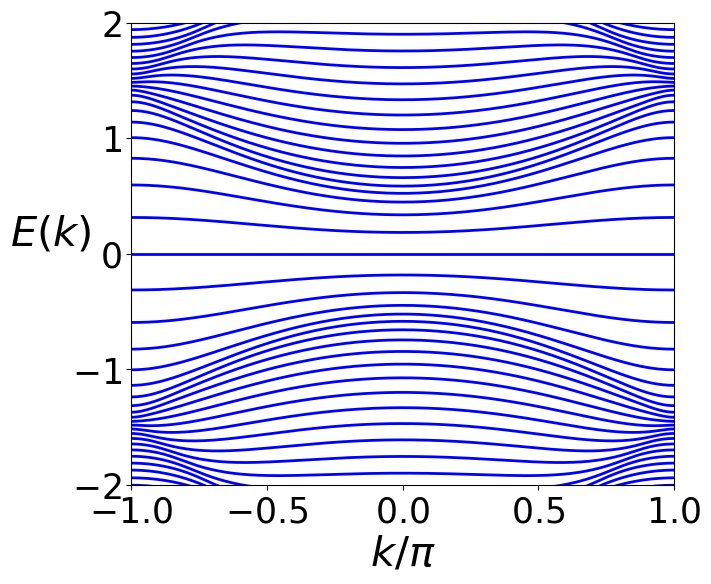}}
\caption{(a), (b) The band structure around the flat band at \(E=0\), obtained by diagonalizing~\eqref{eq:2D_(1_1)_k} and \eqref{eq:(3_1)_ham_square_k}, respectively.
The potential~\eqref{eq:arctan_pot} parameters are \(\lambda = 4\)(a) and \(\lambda = 3\)(b), \(\xi = 5\), and system size \(L = 201\) along \(z\).
}
\label{fig:frac_flat}
\end{figure}

We consider the following trigonometric potential in Hamiltonians~\eqref{eq:2D_(1_1)_k} and~\eqref{eq:(3_1)_ham_square_k},
\begin{align}
  \label{eq:arctan_pot}
  V(z) = \lambda \arctan\left(\frac{z}{\xi}\right),
\end{align}
with \(\xi = 5\), for two different directions \((1, 1)\) with \( z = n_1 + n_2\) and \((3, 1)\) with \(z = 3n_1 + n_2\) respectively.
Figure~\ref{fig:frac_flat} shows the spectrum computed numerically for the potential strengths $\lambda = 4$ and $\lambda = 3$ for directions $(1,1)$ and $(3,1)$, respectively;
the system size is $L = 201$ along the $z$ direction.
Both spectra are symmetric with respect to \(E = 0\) and feature a \(E=0\) flat band.

\begin{figure}
\subfigure[]{\includegraphics[width = 0.465\columnwidth]{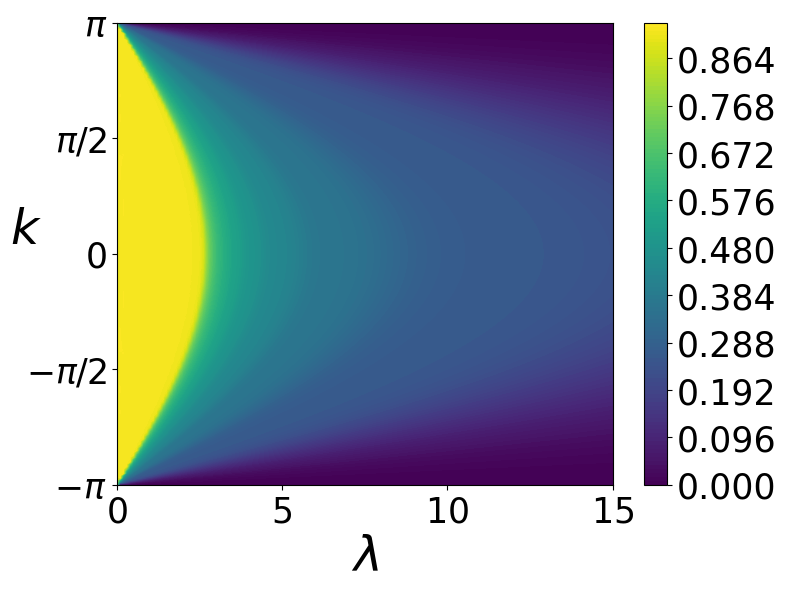}}
\subfigure[]{\includegraphics[width = 0.45\columnwidth]{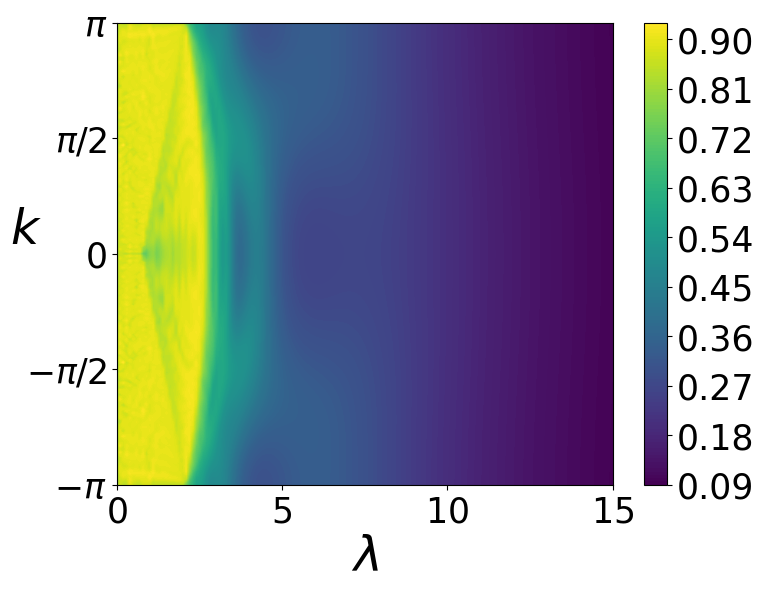}}
\caption{Rescaled participation entropy \(\mathcal{S}(k,\lambda)/\ln(L)\) of the \(E = 0\) flat-band eigenstates as a function of \(k, \lambda\) for the potential~\eqref{eq:arctan_pot}. System size \(L = 201\) along the potential \(z\), \(\xi = 5\). Potential direction \((1, 1): z = n_1 + n_2\) (a) and \((3, 1): z = 3 n_1 + n_2\) (b).}
\label{fig:pr_arctan_flat}
\end{figure}

\begin{figure}
  \subfigure[]{\includegraphics[width = 0.23\textwidth]{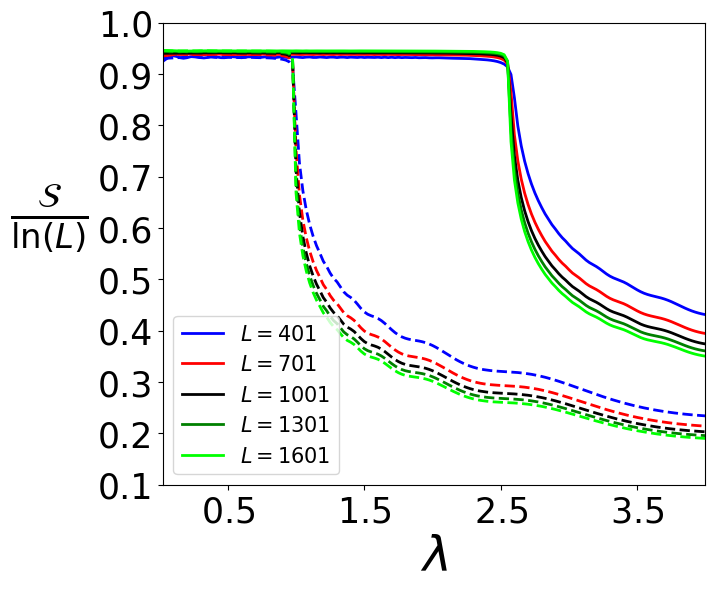}}
  \subfigure[]{\includegraphics[width = 0.23\textwidth]{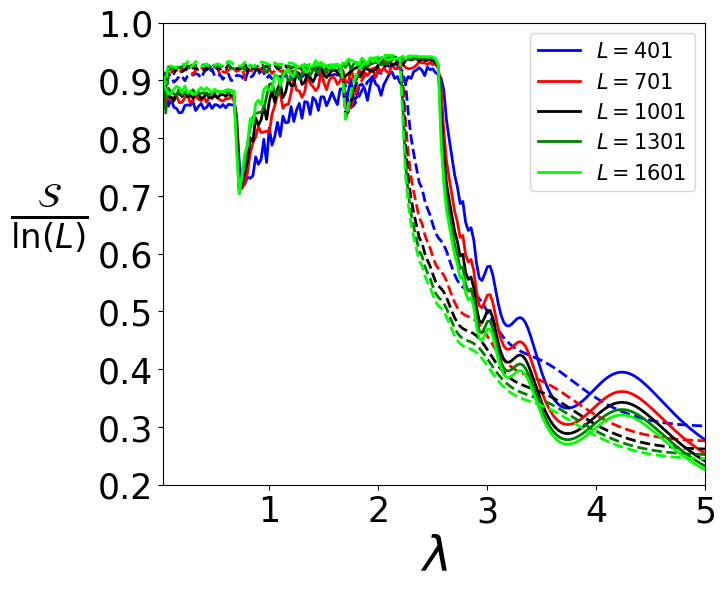}}
  \caption{
    Participation entropy in units of \(\ln(L)\) as a function of the potential strength \(\lambda\) for the potential~\eqref{eq:arctan_pot} with \(\xi = 5\), for different system sizes \(L\) along the $z$ direction.
    Potential directions (a) \((1, 1), z = n_1 + n_2\) and (b) \((3, 1), z = 3n_1 + n_2\).
    Solid lines correspond to momentum \(k = 0\) while dashed lines correspond to \(k = 3\pi/4\).
  }
  \label{fig:pr_scaling_arctan}
\end{figure}

Similarly to the case of the Aubry-Andr\'e potential, the participation entropy \eqref{eq:ipr_ent} for flat-band eigenstates for different momenta \(k\) reveals a localization-delocalization transition along the $z$ direction as shown in Fig.~\ref{fig:pr_arctan_flat}.
For the Aubry-Andr\'e potential, the transition was also revealed by the analytical argument, however,  no such argument is known for a generic bounded potential, like Eq.~\eqref{eq:arctan_pot}.
Therefore, to confirm the transition, we analyzed the finite-size scaling with the increase of the system size $L$ along the $z$ direction.
The results are shown in Fig.~\ref{fig:pr_scaling_arctan} for the two orientations of the potential, two representative momenta \(k=0,3\pi/4\), and multiple system sizes \(L\) along the \(z\) direction.
Participation entropy curves overlap on top of each other or increase to saturation for small values of \(\lambda\) with the system size, while the curves decrease with system size for large values of \(\lambda\).
This strongly supports a localization-delocalization transition as a function of \(k\) and \(\lambda\) for the arctan potential, similarly to the AA case.
The presence of localized and delocalized eigenstates in the flat band for the AA~\eqref{eq:1D_sin_aa} and arctan~\eqref{eq:arctan_pot} potentials suggest that this might be a generic feature of the models~\eqref{eq:main_hamil} with bounded potentials.

\begin{figure}
\includegraphics[width = 0.45\textwidth]{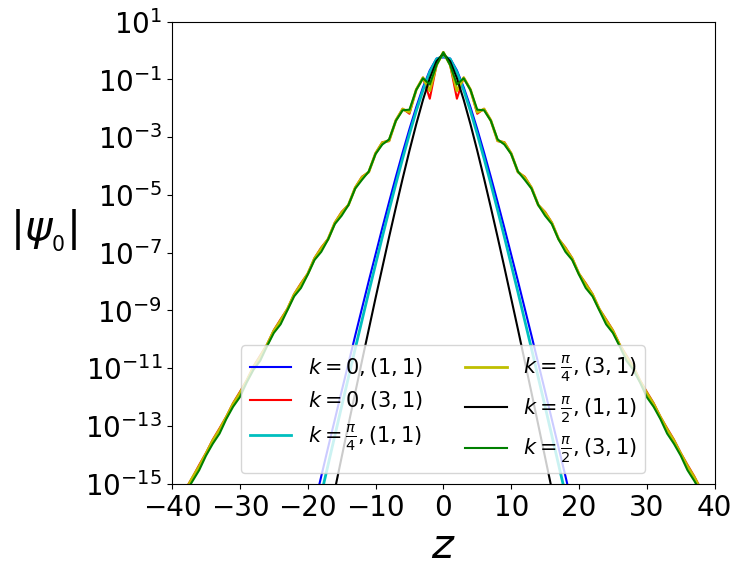}
\caption{
Profiles of the flat-band eigenstates as a function of \(z\) for different momentums and different orientations of the potential~\eqref{eq:arctan_pot} for \(\lambda = 15\). For identical parameters, profiles are more localized for larger momenta $k$.
The direction \((3,1)\) leads to larger localization volume than direction \((1,1)\) for the same set of parameters.
}
\label{fig:arctan_pro}
\end{figure}

Figure~\ref{fig:arctan_pro} shows profiles of the flat band eigenstates for two potential directions \((1, 1), (3, 1)\) and several momenta.
The profiles are more localized for the \((1, 1)\) direction as compared to the \((3, 1)\) direction, as well as for larger momenta \(k\).

\subsection{Unbounded polynomial potential}
\label{sec:poly_flat}

The two examples provided above correspond to bounded potentials, i.e., \(|V(z)| < \infty\).
Here we consider an example of an unbounded polynomial potential:
\begin{align}
  \label{eq:poly_pot}
  V(z) = \lambda
  \begin{cases}
    z^2 & z > 0 \\
    0   & z = 0 \\
    -z^2 & z < 0
  \end{cases}
\end{align}
in Hamiltonians~\eqref{eq:2D_(1_1)_k} for \((1, 1): z = n_1 + n_2\) and~\eqref{eq:(3_1)_ham_square_k} for \((3, 1): z = 3 n_1 + n_2\).

\begin{figure}
\subfigure[]{\includegraphics[width = 0.45\columnwidth]{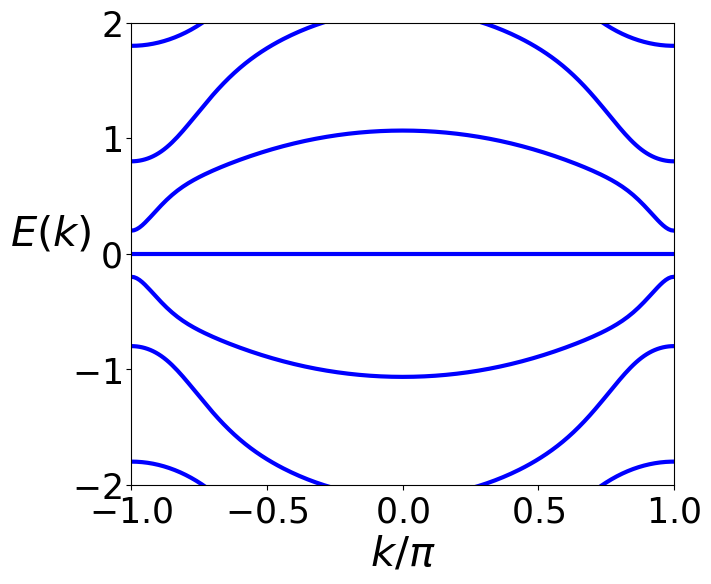}}
\subfigure[]{\includegraphics[width = 0.45\columnwidth]{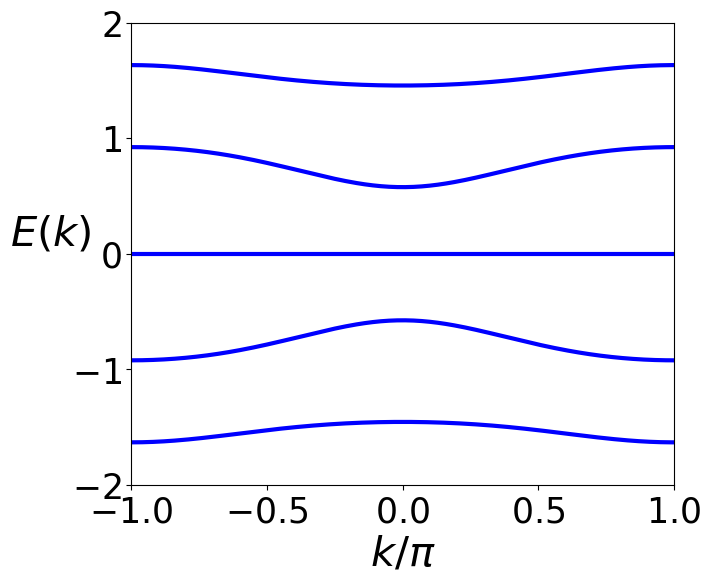}}
\caption{
Part of the band structure contains the flat band \(E = 0\).
The band structure is obtained by diagonalizing~\eqref{eq:2D_(1_1)_k} for (a) and \eqref{eq:(3_1)_ham_square_k} for (b) for system size \(L=201\) along \(z\) for potential~\eqref{eq:poly_pot} with potential strength \(\lambda = 0.2\).
}
\label{fig:poly_flat}
\end{figure}

Figure~\ref{fig:poly_flat} shows the numerically computed spectrum for system size \(201\) along the \(z\) direction and potential strength \(\lambda=0.2\).
The spectrum is symmetric with respect to \(E = 0\)
and there is a \(E = 0\) flat band.
Interestingly, we do not observe an all-bands-flat spectrum in this case in contrast to the linear Wannier-Stark case~\cite{mallick2021wannier}, although the potential is also unbounded in this case.

\begin{figure}
\includegraphics[width = 0.45\textwidth]{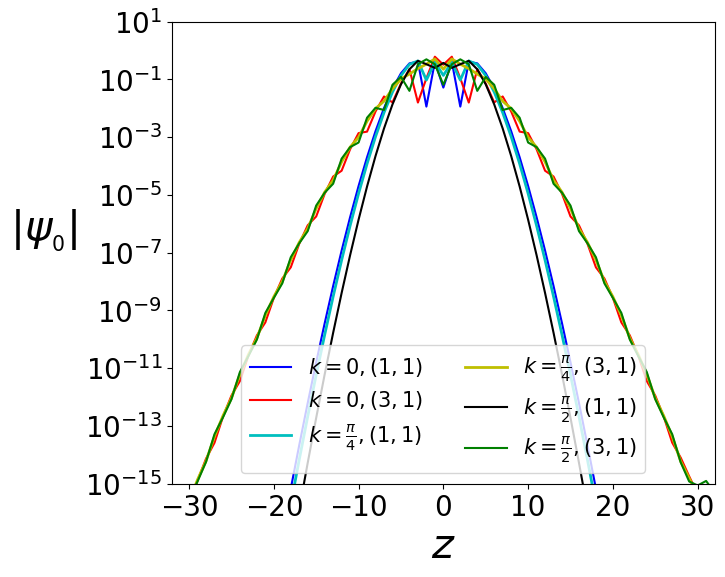}
\caption{
Profiles of the flat-band eigenstates as a function of \(z\) for different momenta and different orientations of the unbounded potential~\eqref{eq:poly_pot} at fixed \(\lambda = 0.2\).
The direction \((3,1)\) leads to larger localization volume than direction \((1,1)\) for the same set of parameters.
}
\label{fig:z2_pro}
\end{figure}

Unlike the bounded potentials, variation of the potential strength \(\lambda\) does not change localization properties of the flat-band eigenstates: they are always localized.
Still localization of the eigenstates depends weakly on the momentum \(k\) and is more pronounced for different directions of the potential as shown in Fig.~\ref{fig:z2_pro}. The asymptotic localization properties in 2D
(3D) Bravais lattices for arbitrary polynomial potential can be
estimated using the approximate analytical treatment reported
in Sec.~II B of Ref.~\cite{mallick2023intermediate}, which suggests a superexponential
decay along the \(z\) axis in our setting.

\section{Conclusion}
\label{sec:concl}
We considered square lattice with tight-binding Hamiltonians and anisotropic on-site potentials that vary along one direction only.
If such a Hamiltonian is anti-symmetric under simultaneous reflection and time-reversal a flat band emerges at \(E=0\) for odd lattice sizes along the potential direction.
This flat-band construction extends to other Bravais lattices (see Appendix~\ref{app:general_lattice}), and can be thought of as a generalization of the Wannier-Stark flat bands.
The existence of the \(E=0\) flat band forbids equipotential hopping, but is otherwise robust to the presence of longer-range hoppings under some mild constraints.

We conjectured the absence of CLS for the \(E=0\) flat band considered.
The localization properties of the flat-band eigenstates were analyzed in the quasimomenta (defined along the equipotential lattice direction) basis, which implies an extended wave packet along the equipotential lattice direction.
For the two bounded potentials, we observe a localization transition in the flat-band eigenstates along the direction of the change of the potential with the increase of the potential strength,
while no such transition is observed for unbounded potentials.
For the quasiperiodic Aubry-Andr\'e potential, this transition is described analytically using a duality transformation.
Understanding whether such transformation exists for other potentials and its identification for other bounded potentials is an open problem.

The Ref.~\cite{mallick2022antipt} introduced an anti-\(\mpt\) symmetry which ensures an \(E=0\) flat band in non-Bravais lattices. The parity operator associated with the anti-\(\mpt\) symmetry acts uniformly at each unit cell and cannot be defined for tight-binding Hamiltonians on Bravais lattices.
The parity operator introduced in this paper is different from that used in Ref.~\cite{mallick2022antipt}: it distinguishes odd and even lattice sites along the potential and is defined for Bravais lattices.

For Bravais lattices, tight-binding Hamiltonians in a dc field, i.e., linear potential, and in the absence of equipotential hopping, all the bands in the spectrum are flat~\cite{mallick2021wannier}.
We observed that for other types of unbounded potentials, the Hamiltonian does not support an all-bands-flat structure, in general.
Presence of a flat band combined  with dispersive bands can have special importance for nontrivial topology and geometry induced exotic phenomena~\cite{chen2014the, julku2016geometric}.

A possible setting to implement Hamiltonians considered
in our works could be single magnon excitations in quantum
XX spins on higher dimensional lattices. A single magnon
Hamiltonian would map onto a single particle tight-binding
problem. A longitudinal magnetic field with a specific spatial
profile applied to a single magnon in the spin system would
then correspond to the anisotropic potential. One experimental
platform is cold atomic devices based on Rydberg atoms,
e.g., $^{87}$Rb. The ground and excited states of the Rydberg atom are equivalent, respectively, to the spin-down (or absence of particle) and spin-up (or presence of particle) states
\cite{jaksch1998cold,ferracini2022realization,vylegzhanin2023excitation}. The detuning parameter provides extra potential energy to the excited Rydberg state compared to its ground state. Anisotropically varying the detuning parameter for the array
of $^{87}$Rb Rydberg atoms could mimic the desired
longitudinal magnetic Z potential for the XX spins \cite{jaksch1998cold,ferracini2022realization,vylegzhanin2023excitation}. A linear
anisotropic potential along the diagonal direction of a square
lattice using spatially resolved microwave spectroscopy has been implemented recently \cite{adler2024observation}. By changing the Helmholtz coil configuration, we might change the linear potential to desired anisotropic bounded or unbounded potentials \cite{adler2024observation}. A superconducting processor was also used to implement linear gradient potential \cite{wang2025exploring}, which may be modified to realize arbitrary potentials.

\begin{acknowledgments}
We are grateful to Sergej Flach for stimulating discussions. The authors acknowledge the financial support from the Institute for Basic Science (IBS) in the Republic of Korea through Project No.~IBS-R024-D1. A.M.~acknowledges financial support from the National Science Centre, Poland (Grant No.~2021/03/Y/ST2/00186) within the QuantERA II Programme, which has received funding from the European Union's Horizon 2020 research and innovation program under Grant Agreement No.~101017733.
\end{acknowledgments}

\appendix

\section{\(E=0\) flat bands for generic \(d\)-dimensional Bravais lattices with anisotropic  potentials}
\label{app:general_lattice}

We consider here a generalization of the case of the square lattice from Sec.~\ref{sec:model} to the case of a \(d\)-dimensional Bravais lattice.
We use some of the definitions from our previous work on Wannier-Stark flat bands~\cite{mallick2021wannier}.
The lattice sites \(\vec{n} = \sum_{j} n_j \vec{a}_j\) are indexed by a set of integers \(\{n_j\}\) and unit cell basis vectors \(\vec{a}_j\) need not be orthonormal.
We consider an anisotropic potential that varies along a lattice direction \(\vmE\) only and is constant in the perpendicular lattice directions.
We assume that the direction \(\vmE\) is commensurate as defined in Ref.~\cite{mallick2021wannier}: one can choose \((d-1)\) linearly independent lattice vectors perpendicular to \(\vmE\).
Under these conditions, the potential can be expressed as a function of the lattice coordinate \(z = \vmE \cdot \vec{n}/\mF\) only, where the constant \(\mF\) ensures that \(z\) is integer and takes all possible values from \(\mathbb{Z}\)~\cite{mallick2021wannier, mallick2022antipt}.
The tight-binding Hamiltonian reads
\begin{align}
  \mh = \sum_{\vec{n}} &\bigg[V(z) \ketbra{\vec{n}}{\vec{n}} - \sum_{\vec{l}} t(\vec{l}) \ketbra{\vec{n}}{\vec{n} + \vec{l}} \bigg].
  \label{eq:gen_ham_app}
\end{align}
The commensurate potential by definition implies the existence of such \((d-1)\) perpendicular vectors which are lattice vectors with different lattice spacing compared to one in the unit of \(\{\vec{a}_j\}\) for the original unit cell coordinate \(\{n_j\}\). 
We define the rotated coordinates along chosen \((d-1)\)-directions perpendicular to \(z\)~\cite{mallick2021wannier}:
\begin{align}
  w_s = \vmE^\perp_s \cdot \vn,~s = 2, 3, \ldots, d\;. 
\end{align}
The coordinates \(w_s\) are not \(z\) independent, in general, however, they can be expressed as linear combinations of coordinate \(z\) and \(z\)-independent \((d-1)\)-dimensional vector \(\vec{\eta}\).
Each of the components \(\eta_i\) is integer and a linear function of \(\{n_j\}\)---cf.~Eqs.~\eqref{eq:coordinate_def_11} and~\eqref{eq:coordinate_def_31} in the main text for special cases,
and Refs.~\cite{mallick2021wannier, mallick2022antipt} for the general form of \(\vec{\eta}\).
In this new coordinate \((z, \vec{\eta})\), the Hamiltonian~\eqref{eq:gen_ham_app} reads
\begin{align}
  \mh = \sum_{z, \vec{\eta}} &\bigg[V(z) \ketbra{z, \vec{\eta}}{z, \vec{\eta}}- \sum_{\vec{l}} t(l_z, \vec{\epsilon}) \ketbra{z, \vec{\eta}}{z + l_z, \vec{\eta} + \vec{\epsilon}} \bigg],
  \label{eq:gen_ham_app_rotated}
\end{align}
where \(l_z = \vmE \cdot \vec{l}/\mF\), \(\vec{\epsilon}\) is the hopping vector along \(\vec{\eta}\) for a fixed \(z\) and is a linear function of \(\vec{l}\).
Importantly, the Hamiltonian is translationally invariant along the \(\vec{\eta}\) directions.
Therefore, it is diagonalizable with respect to lattice momentum \(\vec{k}\) conjugate to \(\vec{\eta}\):
\begin{gather}
  \mh = \sum_{\vec{k}} \mh(\vec{k}) \otimes \ketbra{\vec{k}}{\vec{k}}, \notag \\ 
  \mh(\vec{k}) = \sum_{z} \bigg[V(z) \ketbra{z}{z}- \sum_{\vec{l}} t(l_z, \vec{\epsilon}) e^{i \vec{k} \cdot \vec{\epsilon}} \ketbra{z}{z + l_z} \bigg]\;.
  \label{eq:gen_ham_app_rotated}
\end{gather}
We define a parity operator which reflects a lattice point with respect to a reference lattice point \(\vec{n} = \vec{0}\):
\begin{align}
  \mcp = \sum_{\vec{n}} e^{i z \pi} \ketbra{-\vec{n}}{\vec{n}}\;.
\end{align}
We consider real and antisymmetric potentials:
\begin{gather*}
  V(z > 0) = -V(z < 0),\qquad V(z = 0) = 0 \;.
\end{gather*}
Therefore, the total Hamiltonian as well as the \(\vk\)-dependent Hamiltonians are antisymmetric,
\begin{gather*}
  (\mct \cdot \mcp) \cdot \mh  = - \mh \cdot  (\mct \cdot \mcp), \\
  \mel{\vk}{\mct \cdot \mcp}{\vk} \cdot \mh(\vk)  = - \mh(\vk) \cdot \mel{\vk}{\mct \cdot \mcp}{\vk},
\end{gather*}
subject to the condition that 
\begin{align}
   & l_z = \vmE \cdot \vec{l}/\mF = 2m+1,\qquad m\in\mathbb{Z}\;.
  \label{appeq:hopp_cond}
\end{align}
This condition prohibits equipotential hoppings.
Additionally, it implies that specific directions \(\vmE\) and long-range hopping do not support an \(E = 0\) flat band.
For examples on a square lattice:
(i) potential change along \(\vmE/\mF = (1,1)\) implies \(z = n_1 + n_2\).
If we include the long-range diagonal hoppings, \(\vl = \pm(1, 1) \implies l_z = \pm 2\) violates~\eqref{appeq:hopp_cond};
(ii) a tilted potential along \(\vmE/\mF = (2,1)\) implies \(z = 2 n_1 + n_2\) and the
nearest-neighbor hoppings \(\vl = (\pm 1, 0)\) violates~\eqref{appeq:hopp_cond}.

The antisymmetry of the Hamiltonian \(\mh(\vk)\) implies that the eigenvalues come in pairs \((E(\vk), -E(\vk))\) with corresponding eigenvectors \((|\psi(\vk)\rangle\), \(\langle\vk|\mct \cdot \mcp|\vk\rangle | \psi(\vec{k})\rangle)\) for each \(\vk\).
Together with a center of reflection \(z = 0\), we have an odd number of degrees of freedom for a fixed momentum, therefore the block Hamiltonian \(\mh(\vec{k})\) contains an odd number of independent eigenvectors.
Therefore, there is an eigenvalue which has to be its own negation: \(E(\vk) = -E(\vk)\), and therefore there is an eigenvalue \(E(\vk) = 0\) for all \(\vk\), i.e., a flat band.

Note that only \(\mcp\) is enough to flip the sign of the potential.
But we use extra \(\mct\).
Individually, \(\mct\) or \(\mcp\) flips the sign of momentum \(\vk\), but together they do not.
This is the reason that the antisymmetry of the whole Hamiltonian transfers to the antisymmetry of block Hamiltonians \(\mh(\vk)\) for every momentum \(\vk\)---which is necessary for the existential proof of the flatband.

\subsection{Impossibility of the anti-\(\mcp \mct\) symmetric flat band in triangular lattice with nearest-neighbor hopping}
\label{app:triangular}

In the main text, Sec.~\ref{sec:tilted_square}, we provided examples of square lattice Hamiltonians with potential along the main diagonal and direction (3,1).
Here we show that for the nearest-neighbor hopping, the triangular lattice geometry does not obey the antisymmetry condition~\eqref{appeq:hopp_cond}.
For the triangular lattice, if we align one of the unit cell basis vectors \(\vec{a}_1\) along a Cartesian axis, the other one will be titled by angle \(\pi/3\): \(\vec{a}_2 \cdot \vec{a}_1 = \cos(\pi/3)\). 
Therefore, in the orthonormal Cartesian basis \((\hat{e}_1 = \vec{a}_1, \hat{e}_2)\), a lattice vector reads
\begin{align}
  \vec{n} = n_1 \vec{a}_1 + n_2 \vec{a}_2 = (n_1 + n_2/2)\hat{e}_1 + (\sqrt{3}n_2/2) \hat{e}_2\;.
\end{align}
A nearest-neighbor Hamiltonian with a potential reads
\begin{gather}
  \mh = \sum_{n_1, n_2} \bigg[V(z) \ketbra{n_1, n_2}{n_1, n_2}
  - \sum_{p = \pm 1} \ketbra{n_1, n_2}{n_1 + p, n_2} \notag\\ + \ketbra{n_1, n_2}{n_1, n_2 + p}  
  + \ketbra{n_1, n_2}{n_1 + p, n_2 - p}\bigg]\;.
  \label{eq:triang_ham}
\end{gather}
Here the coordinate \(z\) is taken as a linear combination of \(n_1\) and \(n_2\),
\begin{align}
  z = \alpha_1 n_1 + \alpha_2 n_2,
\end{align}
with coprime coefficients \(\alpha_1, \alpha_2\)---as ensured by the proper choice of \(\mF\) depending on \(\vmE\)~\cite{mallick2021wannier}. 
Therefore, for the hopping present in Hamiltonian~\eqref{eq:triang_ham}:
\begin{align}
  z \mapsto z - l_z,~ l_z \in \{\pm \alpha_1, \pm \alpha_2, \pm(\alpha_1 - \alpha_2)\}\;.
\end{align}
For the validity of the antisymmetry condition~\eqref{appeq:hopp_cond}, \(\alpha_1\), \(\alpha_2\) and \(\alpha_1 - \alpha_2\) have to be odd numbers simultaneously, which is impossible.

\section{Duality transformation for the quasiperiodic potential}
\label{app:transition_deriv}

\begin{figure}
\centering
\subfigure[]{\includegraphics[width = 0.23\textwidth]{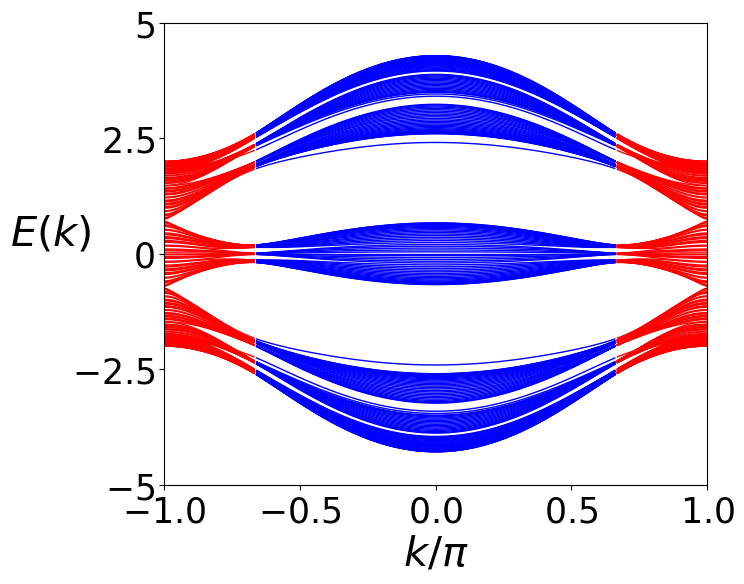}}
\subfigure[]{\includegraphics[width = 0.23\textwidth]{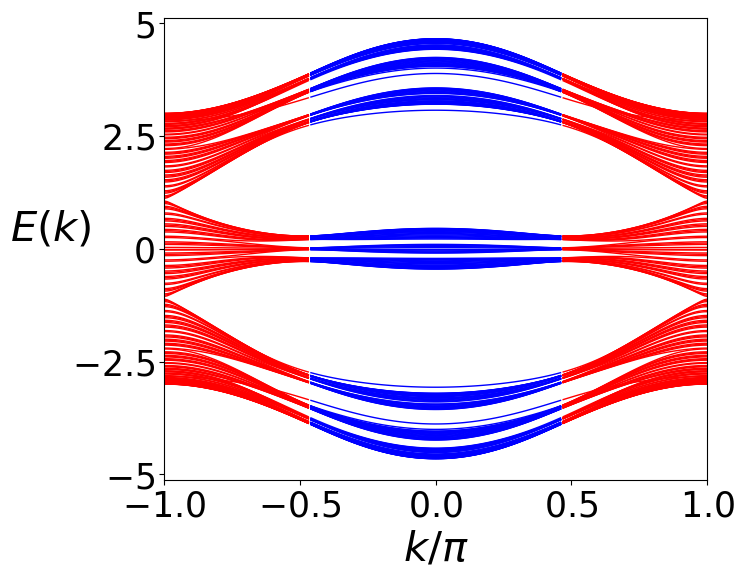}}
\caption{Eigenvalues as a function of momentum for Hamiltonian~\eqref{eq:2D_(1_1)_k} for potential~\eqref{eq:1D_sin_aa} with (a) \(\lambda = 2\) and (b) \(\lambda = 3\).
Red (blue) marks the region with localized (delocalized) eigenfunctions along the \(z\) direction.
The color marked is based on Eq.~\eqref{eq:app_kc}.
Note that increasing \(\lambda\) shrinks the delocalized regimes.}
\label{fig:app_mob_edge}
\end{figure}

The eigenproblem for the Hamiltonian~\eqref{eq:2D_(1_1)_k} with AA potential~\eqref{eq:1D_sin_aa} at each momentum \(k\) reads
\begin{align}
  \label{eq:eigen_2d_AA}
  & E(k) \psi_z(k) = \lambda \sin(\beta z)  \psi_z(k) \notag\\
  & - 2 \cos\left(\frac{k}{2}\right) \left[e^{\frac{ik}{2}}\psi_{z+1}(k) + e^{-\frac{ik}{2}}\psi_{z-1}(k) \right],
\end{align}
with \(\ket{\psi(k)} = \sum_z \psi_z(k) \ket{z}\).
This eigenproblem is equivalent to that of the Hamiltonian in Eq.~(55) of the Ref.~\cite{rossignolo2019localization} up to the additional phase factors \(e^{\pm\frac{ik}{2}}\) in the hopping parameters.
Inspired by the duality transformations used in Refs.~\cite{danieli2016advances,rossignolo2019localization}, we use a modified Fourier transformation connecting \(z\) space to \(m\) space:
\begin{align}
  \label{eq:app_fourier}
  \psi_z(k) =  \frac{1}{\sqrt{L}} e^{i\gamma z}\sum_{m} g_m e^{2 \pi i m \beta z} (i)^m e^{\frac{imk}{2}}\;.
\end{align}
The value of parameter \(\gamma\) is independent of \(m\) and \(\beta\) and is to be determined later. \(L\) is the size of the lattice along the \(z\) direction.
Plugging the modified Fourier transform into Eq.~\eqref{eq:eigen_2d_AA} we arrive at
\begin{gather}
  E(k) g_m =  - \frac{\lambda}{2} \left[ g_{m-1} e^{-\frac{ik}{2}} + g_{m+1} e^{\frac{ik}{2}}\right] \notag \\
  - 4 \cos\left(\frac{k}{2}\right) \cos\left(2 \pi m \beta + \gamma + k/2\right) g_m\;.
\end{gather}
Setting \(\gamma + k/2 = \pi/2\), so \(\cos\left(2 \pi m \beta + \gamma + k/2\right)\) \(= -\sin(2 \pi  m \beta)\), we arrive at the dual problem:
\begin{gather}
  E(k) g_m =  - \frac{\lambda}{2} \left[g_{m-1} e^{-\frac{ik}{2}} + g_{m+1} e^{\frac{ik}{2}}\right] \notag \\
  + 4 \cos\left(\frac{k}{2}\right) \sin(2 \pi m\beta) g_m\;.
\end{gather}
This eigenproblem is equivalent to Eq.~\eqref{eq:eigen_2d_AA} up to a swap of \(\lambda\) and \(4 \cos\left(\frac{k}{2}\right)\).
Therefore, the Hamiltonian~\eqref{eq:2D_(1_1)_k} with potential~\eqref{eq:1D_sin_aa} has a duality transformation similar to that of the Aubry-Andr\'e model, but for every individual momentum \(k\).
We expect a localization to delocalization transition happen at a self-dual point:
\begin{align}
  \label{eq:app_aa_trans}
  \lambda = 4\cos\left(\frac{k}{2}\right)\in [0, 4],~~k \in [-\pi, \pi] \;.
\end{align}
This expression gives a localization-delocalization transition curve in the \((\lambda, k)\) space and we can determine the boundary in \(k\):
\begin{align}
  E = E(k_c);~ k_c = 2 \arccos\left(\frac{\lambda}{4}\right) \in  [0, \pi]\;.
  \label{eq:app_kc}
\end{align}
For a fixed \(\lambda \leq 4\), all eigenstates, including the \(E = 0\) flat-band eigenstates, with energies \(E \in\{E(k): \abs{k} < k_c\}\) delocalize along the \(z\) direction, while all eigenstates with \(E \in\{E(k): \abs{k} > k_c\}\) are localized along the \(z\) direction. Figure \ref{fig:app_mob_edge} depicts the mobility edges in the \(k\) space at fixed \(\lambda\).

\bibliography{flatband, mbl, general, ergodicity}

\end{document}